\documentclass[%
    preprint,
    superscriptaddress,
 amsmath,amssymb,
pra,
]{revtex4-2}

\usepackage[utf8]{inputenc}
\usepackage[utf8]{inputenc}%
\usepackage[dvipsnames]{xcolor}
\usepackage{graphicx}%
\usepackage{soul}
\usepackage{todonotes}

\let\vec\boldsymbol

\usepackage{hyperref}%
\hypersetup{
    colorlinks=true,
    linkcolor=blue,
    filecolor=magenta,      
    urlcolor=cyan,
    citecolor = cyan,
    pdftitle={Overleaf Example},
    pdfpagemode=FullScreen,
    }
    
\graphicspath{{Figures/}}

\begin{document}

\title{Time-frequency analysis of nonlinear Compton scattering via joint probability distributions}

\newcommand{\HIJ}{Helmholtz Institute Jena, Fröbelstieg 3, 07743 Jena, Germany}
\newcommand{\GSI}{GSI Helmholtzzentrum für Schwerionenforschung GmbH, Planckstraße 1, 64291 Darmstadt, Germany}
\newcommand{\IOQ}{Institute of Optics and Quantum Electronics, Friedrich-Schiller-Universität, Max-Wien-Platz 1, 07743 Jena, Germany}

\author{Nikita Larin}
\email{nikita.larin@uni-jena.de}
\affiliation{\HIJ}
\affiliation{\GSI}
\affiliation{\IOQ}

\author{Daniel Seipt}
\affiliation{\HIJ}
\affiliation{\GSI}
\affiliation{\IOQ}

\date{\today}

\begin{abstract}

The interaction of charged particles with an intense laser pulse gives rise to a number of characteristic spectral features of emitted radiation, including the generation of harmonics, spectral broadening due to the phase-dependent ponderomotive red shift, and the emergence of intricate sub-harmonic structures. These effects are accumulated over the course of the interaction with the electromagnetic field and are therefore inherently nonlocal in nature. For a deeper understanding of strong-field quantum electrodynamics (SFQED) processes and their practical applications, it is desirable to employ tools that enable simultaneous analysis in the time and energy domains. In time-frequency analysis, such tools are provided by joint distributions (JDs). In this work, we demonstrate how a JD can be devised within the SFQED framework. Specifically, we focus on constructing a non-negative JD, which allows for a clear probabilistic interpretation. We study the properties of the proposed distribution and test its utility by applying it to the nonlinear Compton scattering in complex laser pulse configurations with carrier-envelope phase and variable polarization.

\end{abstract}

\maketitle

\section{Introduction}

Strong-field quantum electrodynamics (SFQED) describes the interaction of charged particles and photons with electromagnetic fields of extreme intensity, such as those produced by modern high-power laser systems \cite{Danson2019, Yoon2021, Maksimchuk2025}. In this regime, the interaction between electrons, positrons, and photons with the background field becomes intrinsically nonlinear \cite{DiPiazza2012, Gonoskov2022, Fedotov2023}, giving rise to processes that have no counterpart in perturbative QED. Prominent examples include nonlinear Breit–Wheeler pair production \cite{Reiss1962} and nonlinear Compton scattering \cite{Ritus1985}, as well as higher-order reactions such as trident processes \cite{Ilderton2011, Torgrimsson2020}, laser-induced double Compton scattering \cite{Seipt2012}, and vacuum birefringence \cite{Ahmadiniaz2025}.

The electromagnetic field of an intense laser pulse exhibits a rich and rapidly varying temporal structure, which plays a central role in shaping its interaction with charged particles. In particular, the emission spectra and angular distributions of radiation are strongly influenced by the pulse envelope \cite{King2021} and cycle-scale field variations \cite{Titov2014}. Despite this explicit dependence on the instantaneous properties of the electromagnetic field, standard SFQED treatments are formulated in terms of asymptotic in- and out-states, and the resulting observables are typically constructed as spectral quantities integrated over the full interaction time.

These considerations motivate the development of methods capable of resolving both the temporal and spectral characteristics of emitted radiation and produced particles in strong fields. Such capabilities are becoming increasingly relevant not only for the interpretation of strong-field QED processes, but also in view of recently proposed high-intensity atto- \cite{Zhang:PRL2020, Xin2023} and zeptosecond \cite{Chen2025, Li2025} radiation sources, which allow probing sub-cycle dynamics. Thus, a detailed understanding of the interplay between the emitted energy and the time structure of the driving laser pulse is required.

Time–frequency analysis provides a natural framework for this purpose through the use of joint distributions (JDs), which allow one to simultaneously characterize when radiation is emitted and at which energies \cite{Cohen1989, Cohen1995}. However, generic joint distributions often exhibit oscillatory behavior and may attain negative values due to the interference effects. While such distributions are useful for revealing correlations and sub-cycle dynamics, their sign-indefinite nature prevents a straightforward probabilistic interpretation and complicates their use in applications where positivity is essential.

One of such application is implementation in numerical modeling of SFQED processes. Contemporary simulation frameworks rely on local emission and pair-creation probability rates as input for Monte Carlo algorithms \cite{Ridgers2014, Gonoskov2015, Fedeli2022}. The two standard probability rates are derived within the locally constant \cite{Ritus1985} and locally monochromatic field approximations \cite{Heinzl2020}. These rates are intrinsically non-stationary and depend on the instantaneous properties of the external field. They determine the probability of the particular event to happen at the given moment of time. In this sense, the probability rate itself can be viewed as a particular realization of a positive-definite JD. This observation motivates the construction of non-negative JD beyond the local approximations, thereby offering a new insight into the validity and limitations of commonly used local approximations, which underpin the existence of instantaneous probability rates in the first place.

Alternating in sign joint distributions have already been considered in the context of SFQED. In particular, they have been employed in studies of quantum vacuum effects in strong electromagnetic fields \cite{BialynickiBirula1991, Hebenstreit2011}, as well as in investigations of radiation emitted by a single-electron wave packet interacting with a laser pulse \cite{Peatross2008}. Similarly, in strong-field atomic physics, time–frequency analysis methods, including the Gabor transform \cite{Cohen1995}, have proven useful for studying strong-field phenomena such as high-harmonic generation \cite{Pukhov2003, Kohler2010}.

The paper is organized as follows. In Sec.~\ref{subsec_JD}, we show how an alternating in sign JD can be devised within the $S$-matrix formalism in the Furry picture and discuss its main properties. In Sec.~\ref{subsec_Husimi}, we demonstrate how this JD can be used to construct a positive-definite one by means of the Husimi transform. Its properties are further analyzed and contrasted with existing local approximations in Sec.~\ref{subsec_Husimi_vs_local}. In Sec.~\ref{sec_results}, we assess the capabilities of the Husimi distribution and apply it to the investigation of carrier-envelope phase effects in Sec.~\ref{subsec_CEP} and of pulses with polarization gating in Sec.~\ref{subsec_PG}. Finally, we summarize our findings in Sec.~\ref{sec_summary}. 

In this paper, we adhere to the Heaviside-Lorentz natural system of units ($c = \hbar = \epsilon_0 = 1$, $\alpha = e^2/4\pi$ is the fine structure constant). We employ light-front coordinates, in which the scalar product of any four-vectors become $x\cdot y = x_\mu y^\mu = x^+y^-/2 + x^-y^+/2 - \vec{x}^\bot\vec{y}^\bot$, with $x^\pm=x^0\pm x^3$ and $\vec{x}^\bot=\left(x^1,x^2\right)$. The hat notation stands for the contraction of the corresponding four-vector with the Dirac gamma matrices $\hat{a}=a_\mu\gamma^\mu$.

\section{Theory and Methods}
\label{sec_theory_methods}

Spectral features arising in strong-field quantum electrodynamics (SFQED) are commonly analyzed using asymptotic observables derived from standard $S$-matrix calculations \cite{Gonoskov2022, Fedotov2023}. While this approach provides direct access to energy-resolved quantities, it does not capture the temporal structure of the processes. Below we show that it is possible, within the SFQED framework, to study energy-temporal structure of the process simultaneously, by operating with the quantity that is suitable for a role of joint distribution. We adapt concepts and methods from time–frequency analysis to construct a non-negative joint distribution, enabling a transparent probabilistic interpretation. We further demonstrate that the devised joint distribution reproduces the main features of the emitted spectrum while providing additional insight into its temporal structure.

\subsection{Joint probability distribution in SFQED}
\label{subsec_JD}

\begin{figure}[h]
\centering
\includegraphics[width=0.5\textwidth]{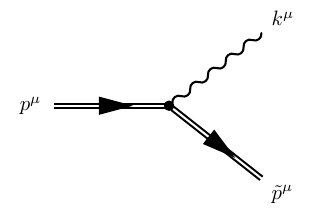}
\caption{Feynman diagram for NCS. Double lines stand for the dressed electron states, the wavy line corresponds to the emitted photon.
    \label{fig_Compton}}
\end{figure}

To be specific and to demonstrate how one can devise a joint distribution for a first order (in the fine structure constant $\alpha$) SFQED process, we consider the nonlinear Compton scattering (NCS) \cite{Fedotov2023} as an illustrative example. By NCS we understand scattering of an initial electron with the four-momentum $p^\mu$ off a classical plane electromagnetic wave with emission of a photon $k^\mu = (k^0,\vec k)$ and subsequent transition to the final state with the four-momentum $\tilde{p}^\mu$ (see Feynman diagram in Fig.~\ref{fig_Compton}). The classical plane electromagnetic wave is characterized by the frequency $\omega$, wave four-vector $\kappa^\mu$ and the normalized four-vector potential ${a_\mu\left(\phi\right)=m\vec{a}_\bot\left(\phi\right)} =m a_0\vec{f}_\bot\left(\phi\right)$. The function $\vec{f}_\bot\left(\phi\right)$ describes laser field profile and $a_0 = |e|E/m\omega$ is the classical nonlinearity parameter \cite{Gonoskov2022}, with $m$, $e$ and $E$ are being the electron mass, charge, and the field strength, respectively.

As a starting point we consider the regularized amplitude for photon emission in a plane wave background \cite{Boca2009, King2021}:
\begin{equation}
{\mathcal{S}}\left(\ell\right) = -ie{\left( {2\pi } \right)^3}\delta^{(3)}_\mathrm{LF}\left( p - k - \tilde{p}\right)\int\limits_{-\infty}^{+\infty} {\mathcal{M}\left(\phi,\ell\right)e^{i\ell\phi} }d\phi,
    \label{Amplitude}
\end{equation}
where $\delta^{(3)}_\mathrm{LF}(p) = \delta(p^-)\delta(\vec{p}^\bot)$ is a light-front delta function. $\mathcal{M}\left(\phi,\ell\right)$ is a complex-valued function, which contains all information about the background field (shape, duration, and polarization) as well as the initial and final particles (their momenta, polarization, and spin):

\begin{align}
\mathcal{M}\left(\phi,\ell\right)=\bar{u}_{\tilde{p}}\left(\Psi\left(\phi\right)\hat{\varepsilon}^* + \frac{\hat{a}\hat{\kappa}\hat{\varepsilon}^*}{2\kappa\cdot\tilde{p}}+\frac{\hat{\varepsilon}^*\hat{\kappa}\hat{a}}{2\kappa\cdot p}-\frac{a^2\kappa\cdot\varepsilon^*}{2\kappa\cdot\tilde{p}\kappa\cdot p }\hat{\kappa}\right)u_pe^{-i\ell\int\limits^\phi\Psi\left(\phi'\right)d\phi'},
\end{align}
with $u_p$ and $\bar{u}_{\tilde{p}}$ are being the incoming and outgoing electron bispinors and $\varepsilon^*_\mu$ is the emitted photon polarization vector. The function $\Psi\left(\phi\right)$ is given by
\begin{align}
    \Psi\left(\phi\right)=1-\frac{1+\left(\vec{a}_\bot\left(\phi\right)+\vec{\rho}_\bot\right)^2}{1+\vec{\rho}_\bot^2},
\end{align}
where $\vec{\rho}_\bot=\left(\frac{\kappa\cdot p}{\kappa\cdot k}\vec{k}_\bot-\vec{p}_\bot\right)/m$ is scaled transverse momentum of the emitted photon.
$\ell$ is an exchanged momentum between the electron and background field \cite{Seipt2017a}. It encodes energy of the emitted photon ($\ell = k\cdot p/\left(\kappa\cdot p - \kappa\cdot k\right) \geq 0$) and together with the laser phase $\phi=\omega x^+$ (with $x^+$ is being a light-front time \cite{Heinzl2001}) constitutes a pair of canonically conjugate variables \cite{seipt_thesis}.

The probability for emitting a photon may be obtained from the amplitude, Eq.~\eqref{Amplitude}, in a standard way \cite{Ilderton2019a} after performing phase integral in \eqref{Amplitude} either analytically \cite{Seipt2016} or numerically \cite{Seipt2011}. Doing so, the information about the temporal structure of the emitted radiation is lost, and we are left with its energy spectrum. To study temporal dynamics of the emission process along with its spectral composition, we choose to preserve the explicit dependence of the amplitude \eqref{Amplitude} on the laser phase. After mod-squaring it, we follow the standard procedure \cite{Ilderton2019a} by multiplying result with the Lorentz invariant light-front phase space element of the final states and performing integration over the electron final momenta with aid of the delta function. 
Throughout this paper we always consider unpolarized quantities, i.e. we average over the initial electron spin and sum over all final spin and polarization states \footnote{The question of existence of the positive-definite fully differential and polarization/spin-resolved joint distribution in constant crossed field was recently addressed in \cite{Montefiori:arXiv2026}}.
The total unpolarized probability for NCS is given by \cite{Fedotov2023}
\begin{align}
    \mathbb P = \int d^2 \rho_\perp d \ell \int d\varphi   \: \frac{d \mathcal P}{d\ell d\varphi d^2\rho_\perp}  = \int d^2\rho_\perp d \ell \int d\varphi   \: \frac{d \mathcal R}{d\ell d^2\rho_\perp}\,,
\end{align}
where we implicitly defined the (quasi-)rate as $\mathcal R = d\mathcal P/ d\varphi$,
\begin{equation}
\frac{d\mathcal{R}\left(\varphi,\ell\right)}{d\ell d^2\rho_\perp} 
= \frac{1}{2}\sum_\mathrm{pol} \frac{\alpha}{4\pi^2m^2} \frac{\ell}{(1+\vec{\rho}_\perp^2+2\eta\ell)^2} \int\limits_{ - \infty }^{ + \infty } {\mathcal{M}\left( {\varphi  + \frac{\theta }{2}},\ell \right){\mathcal{M}^*}\left( {\varphi  - \frac{\theta }{2}},\ell \right)e^{-i\ell\theta}}d\theta.
    \label{Proto_rate}
\end{equation}
Here we introduced the initial electron energy parameter $\eta=\kappa\cdot p/m^2$ and the new variables: the average phase $\varphi=(\phi+\phi')/2$ and the interference window $\theta=\phi'-\phi$ \cite{Fedotov2023} (with $\phi$ and $\phi'$ being the two phases of the amplitude and its conjugate). The integral over $\theta$ accounts for the interference effects that stem from the non-local emission of photon with momentum $k^\mu$ from different parts of the laser pulse. We use curly letters to indicate that the corresponding quantities are not strictly positive definite.

Equation~\eqref{Proto_rate} contains all information about emitted in NCS photon simultaneously in the energy variable (transferred light-front momentum) $\ell$ and laser phase (time variable)  $\varphi$ (as well as $\vec{\rho}_\bot$). Thus, we propose that Eq.~\eqref{Proto_rate} is suitable for the role of a joint distribution for NCS.
However, Eq.~\eqref{Proto_rate} is not necessarily strictly positive. It may attain negative values due to the interference effects. Despite the ambiguous sign of the JD, its marginal distributions, which are obtained by integrating over $\ell$ or $\varphi$, respectively, are strictly non-negative. Integrating \eqref{Proto_rate} over the average phase $\varphi$, we obtain the energy and angular-resolved probability distribution, the spectrum of emitted photons as follows:
\begin{align}
\frac{d\mathbb{P}\left(\ell\right)}{d\ell d^2\rho_\perp } = \int\limits_{-\infty}^{+\infty}\frac{d\mathcal{P}\left(\varphi,\ell\right)}{d\ell  d^2\rho_\perp d\varphi}\:d\varphi\geq0.
    \label{energy_marginal}
\end{align}
In turn, the integration of Eq.~\eqref{Proto_rate} over the transferred momentum $\ell$ 
yields the angular-resolved probability per unit phase, the angular-resolved probability rate,
\begin{align}
    \frac{d\mathbb{P}\left(\varphi\right)}{d\varphi d^2 \rho_\perp } = \int\limits_{0}^{+\infty}\frac{d\mathcal{P}\left(\varphi,\ell\right)}{d\ell d^2 \rho_\perp d\varphi}\:d\ell\geq0.
    \label{phase_marginal}
\end{align}

Performing explicitly the calculations, we obtain for the joint distribution the following expression:
\begin{align}
\frac{d\mathcal{{P}\left(\varphi,\ell\right)}}{d\ell d\varphi}=-\frac{\alpha}{\pi^2}A\int\limits_{-\infty}^{\infty}\left[C_1+BC_2\right]\exp\left[-i\ell\theta\left(1 - \langle \Psi\rangle\right)\right]d\theta \,.
\label{explicit_protorate}
\end{align}
Here and in the following, we will always be dealing with angle-resolved quantities; however, for the sake of brevity, we will omit any explicit mention of the dependence on the transverse momentum in our equations.
The following short-hand notations are introduced,  
\begin{align}
A &= \frac{\ell}{\left(1+\vec{\rho}_\bot^2+2\eta\ell\right)^2},\label{A_func}\\
B &=\frac{1}{2}+\frac{\eta^2\ell^2}{\left(1+\vec{\rho}_\bot^2\right)\left(1+\vec{\rho}_\bot^2+2\eta\ell\right)},\label{B_func}\\
C_1&=-\Psi\left(\varphi+\frac{\theta}{2}\right)\Psi\left(\varphi - \frac{\theta}{2}\right), \\ 
C_2&= \Psi\left(\varphi+\frac{\theta}{2}\right)\vec{a}^2_\bot\left(\varphi-\frac{\theta}{2}\right) 
    + \Psi\left(\varphi-\frac{\theta}{2}\right)\vec{a}_\bot^2\left(\varphi+\frac{\theta}{2}\right) \nonumber\\
    & \qquad -2\vec{a}_\bot\left(\varphi+\frac{\theta}{2}\right)\vec{a}_\bot\left(\varphi-\frac{\theta}{2}\right)\,.\label{Reg_term}
\end{align}
 The angular brackets in Eq.~\eqref{explicit_protorate} denote the floating average \cite{Fedotov2023}:
\begin{align}
    \langle \Psi\rangle &= \frac{1}{\theta}\int\limits_{\varphi-\theta/2}^{\varphi+\theta/2}\Psi\left(\phi\right)d\phi.\label{floating_average}
\end{align}
The functions $C_1$ and $C_2$ in \eqref{Reg_term} contain regularization terms that ensure correct behavior in the limit of vanishing external field \cite{Boca2009,King2021}. 

\begin{figure}[h]
\centering
\includegraphics[width=\textwidth]{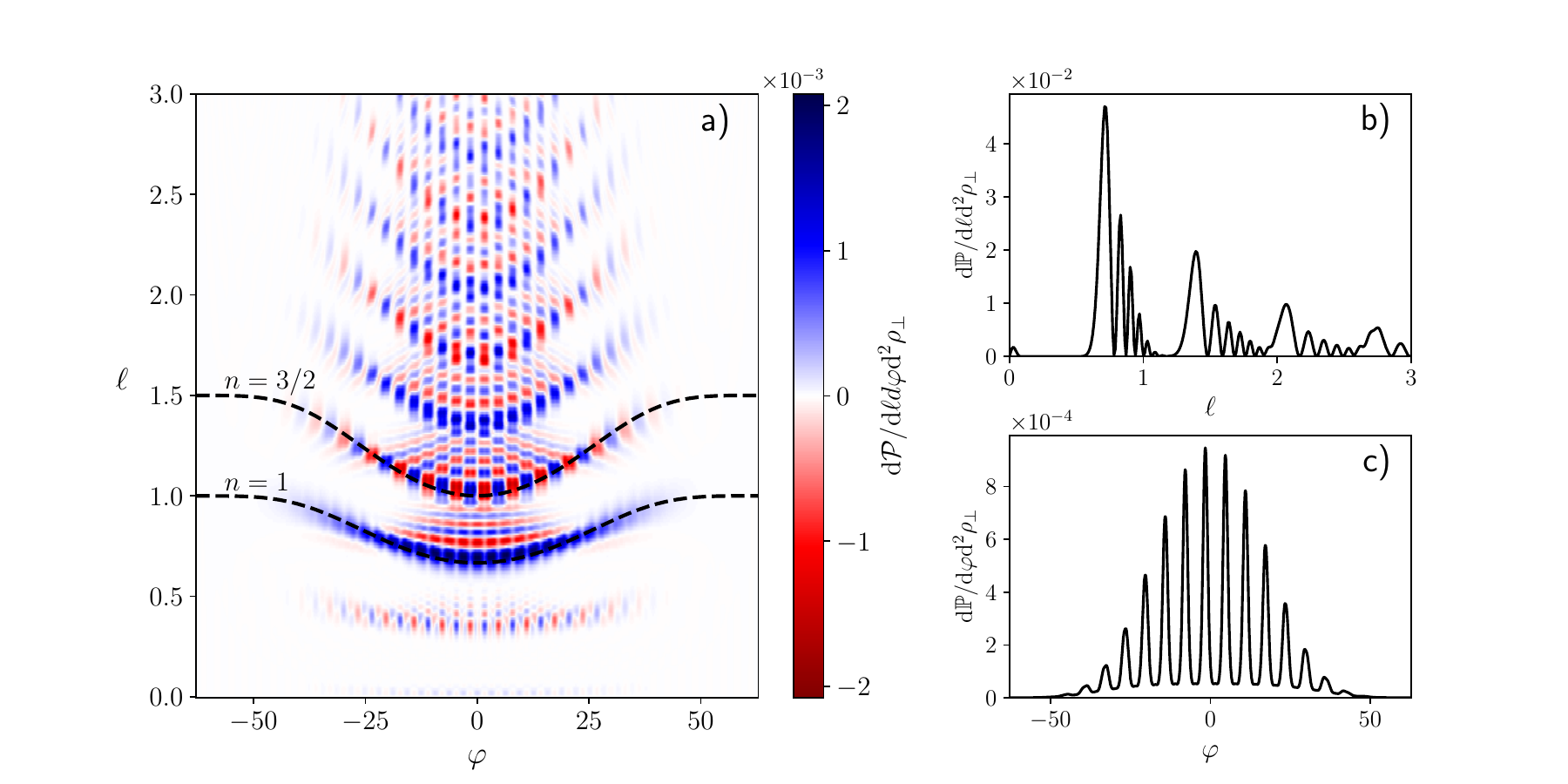}
\caption{a) JD for the NCS and corresponding marginal probabilities with respect to b) the exchanged momentum $\ell$ and c) laser phase $\varphi$. The field potential is chosen as ${{\vec{f}}_ \bot }\left( \phi  \right) =  \cos^2\left(\phi/2N\right)\left( {\sin \phi , \cos \phi } \right)$ with $N = 20$ and the other parameters are: $\eta=0.01$, $a_0 =1$, $\vec{\rho}_\bot=\left(1, 0\right)$. 
    \label{fig_Rate}}
\end{figure}

In Fig.~\ref{fig_Rate} we exhibit the joint distribution for photon emission \eqref{explicit_protorate} and its marginal distributions. We observe that the joint distribution in the $(\varphi,\ell)$-plane alters in sign and has similar features as a Wigner function \cite{Wigner1997}. The negative regions manifest interference effects from the emission of indistinguishable photons, which one accounts when integrates over the interference window $\theta$ in \eqref{explicit_protorate}. This interference leads to the intricate sub-harmonic structure of the spectrum, when \eqref{explicit_protorate} is integrated over the laser phase $\varphi$ (see, Fig.~\ref{fig_Rate}b). Apart from the interference effects, we clearly see the influence of ponderomotive red-shift, which pushes harmonics towards the low-energy region (smaller $\ell$) by different amount depending on the laser phase and results into their broadening. The positions of the harmonics follow $\ell_n(\varphi) = n(1+\vec{\rho}_\bot^2)/(1 + \vec{\rho}_\bot^2 + \vec{a}^2_\mathrm{rms})$ \cite{Seipt2013}, where $\vec{a}_\mathrm{rms}$ is the root-mean-square of the field potential over the carrier period. Noteworthy, that in addition to the integer harmonics, the JD \eqref{explicit_protorate} contains features at half-integer $n$ \cite{Larin2025} with oscillating patterns of positive and negative values. However, the half-integer harmonics average to zero, when the JD is integrated with respect to any of the canonically conjugate variables in calculating the marginals. The latter can be seen to be strictly non-negative, (see Fig.~\ref{fig_Rate}, b and c).

Despite the JD \eqref{explicit_protorate} correctly capturing all the essential non-local features of the photon emission, it has a big disadvantage of not being positively-definite. Thus, we can not assign a probability to the emission of a photon with a certain energy at the given laser phase. This complicates a straightforward interpretation of the result and makes it unsuitable for a direct implementation in numerical codes. 

To conclude this chapter we would like to indicate potential generalizations of our joint distribution.
The key properties in our approach are that (i) the JD in Eq.~\eqref{Proto_rate} is bilinear in the photon emission amplitude and (ii) reproduces the correct marginal distributions. The general theory of the joint distributions suggests that distribution, which possesses both these properties can not be positive-definite everywhere \cite{Wigner1997,Mugur-Schächter1979}. However, it is possible to devise non-negative JDs if one of the listed properties is relaxed. For instance, non-bilinear JDs are studied extensively in time-frequency analysis \cite{Cohen1966, Cohen1980, Finch1984, Cohen1985, Schweizer1986}, but we leave the question of their utility for SFQED for future research.

\subsection{Husimi joint probability distribution}
\label{subsec_Husimi}

To proceed, let us now investigate the role of non-negative JD that are bilinear in amplitude, but do not reproduce exact marginal distributions (i.e., do not reproduce the exact spectrum, when integrated over the time variable) \cite{Cartwright1976, SOTO1981,BERTRAND1983}. Such JDs are usually obtained by the double convolution of a non-positive JD \cite{Cohen1989, Janssen1984,Janssen1985} with a smoothing kernel. The general idea of convolution is to average the negative bits of the original distribution with neighboring positive ones. The result is the coarse-grained JD, whose properties depend strongly on the kernel \cite{Cohen1989, Cohen1995}.

Here we particularly adapt the Husimi transformation \cite{Husimi, Lee1995} of Eq.~\eqref{Proto_rate}, which corresponds to a convolution with two Gaussian kernels in the phase and energy variables, whose dispersions are related via $\sigma_\varphi\sigma_\ell=1/2$. For the phase convolution we use the standard kernel $g_{\sigma_\varphi}(\varphi-\varphi') = e^{ {-\frac{{{{\left( {\varphi - \varphi'} \right)}^2}}}{{2\sigma _\varphi^2}}}} / (\sqrt{2\pi} \sigma_\varphi )$. However, because of the kinematic restriction of $\ell\geq0$, the original JD \eqref{Proto_rate} is not defined in the
whole $(\varphi,\ell)$-plane and we slightly have to modify the kernel in the energy variable $\ell$, which we normalize as follows
\begin{align}
    \int\limits_{0}^{+\infty}\! d\ell'\: g_{\sigma_\ell}(\ell-\ell') = 1\,,
\end{align}
hence $g_{\sigma_\ell}(\ell-\ell') = e^{ {-\frac{{{{\left( {\ell - \ell'} \right)}^2}}}{{2\sigma _\ell^2}}}} / (\sqrt{2\pi} \sigma_\ell I )$ with $I(\ell) = \int\limits_0^{+\infty}\exp(-(\ell - \ell')^2/2\sigma_\ell^2)d\ell'/\sqrt{2\pi\sigma_\ell^2}$.

With this we define the Husimi transformation of the JD in Eq.~\eqref{Proto_rate} as
\begin{align}
    \frac{d\mathbb R_\mathrm{H}}{d\ell} = \frac{{{{d\mathbb{P}_\mathrm{H}}\left( {\varphi ,\ell} \right)}}}{d\ell d\varphi} = \int \limits_{-\infty}^{+\infty} d\varphi' g_{\sigma_\varphi}(\varphi-\varphi')\int\limits_0^{+\infty}d\ell'g_{\sigma_\ell}(\ell-\ell')\frac{d\mathcal{P}\left( {\varphi',\ell'} \right)}{d\ell' d\varphi'}\geq0 \,,
    \label{Husimi_new}
\end{align}
where the subscript "H" stands for Husimi. With this definition, the integrated probability is invariant under the Husimi transformation in the following way:
\begin{align}
    \int\limits_{-\infty}^{+\infty}d\varphi\int\limits_0^{+\infty}d\ell \: \frac{d\mathcal{P}(\varphi,\ell)}{d\ell d\varphi} = \int\limits_{-\infty}^{+\infty}d\varphi\int\limits_0^{+\infty}d\ell \: \frac{d\mathbb{P}_\mathrm{H}(\varphi,\ell)}{d\ell d\varphi}.
\end{align}
Moreover, the corresponding marginals of the Husimi distribution Eq.~\eqref{Husimi_new} become
\begin{align} \label{husimi_phi_marginal}
& \int\limits_{0}^{+\infty}\frac{\mathbb{P}_\mathrm{H}(\varphi, \ell)}{d\ell d\varphi}d\ell=\int\limits_{-\infty}^{+\infty}g_{\sigma_\varphi}(\varphi-\varphi')\frac{\mathbb{P}(\varphi')}{d\varphi'}d\varphi' =: \frac{d\mathbb P_\mathrm{H}(\varphi)}{d\varphi}, \\
\label{husimi_ell_marginal}
&\int\limits_{-\infty}^{+\infty}\frac{\mathbb{P}_\mathrm{H}(\varphi, \ell)}{d\ell d\varphi}d\varphi=\int\limits_{0}^{+\infty}g_{\sigma_\ell}(\ell-\ell')\frac{\mathbb{P}(\ell')}{d\ell'}d\ell'
 =: \frac{d\mathbb P_\mathrm{H}(\ell)}{d\ell},
\end{align}
i.e. the convolutions of the marginals, Eqs.~\eqref{energy_marginal} and \eqref{phase_marginal}, of Eq.~\eqref{explicit_protorate} with their corresponding kernels. The joint distribution, Eq.~\eqref{Husimi_new}, defined in this way is always non-negative, and therefore permits interpretation as a probability for given $\ell$ and $\varphi$. Henceforth, we refer to \eqref{Husimi_new} as a Husimi joint probability distribution (JPD).

The use of the Gaussian functions as a convolution kernel \cite{Cartwright1976, SOTO1981, Janssen1984,Janssen1985} is motivated by the fact that the Wigner joint distribution for a coherent state \cite{Glauber1963} is always non-negative \cite{Hudson1974, SOTO1983}. The condition $\sigma_\varphi\sigma_\ell=1/2$ ensures that the convolution in \eqref{Husimi_new} smears the original distribution over the smallest possible area in the phase-space (i.e., the phase-space cell volume $2\pi\hbar$). The positive-definiteness holds for the convolution with all $\sigma_\varphi \sigma_\ell \geq 1/2$ \cite{Cartwright1976, SOTO1981}, but the maximum contrast of the resulting distribution is achieved for the lower boundary. If one tries to resolve features on scales smaller than the minimal phase-space cell size (i.e., $\sigma_\varphi \sigma_\ell < 1/2$) the result may become negative \cite{Janssen1984}. Previously, the convolution of phase-space joint distributions with wide Gaussians was employed for studying the Schwinger effect in inhomogeneous electromagnetic fields \cite{Kohlfurst2018}.

\begin{figure}[h]
\centering
\includegraphics[width=\textwidth]{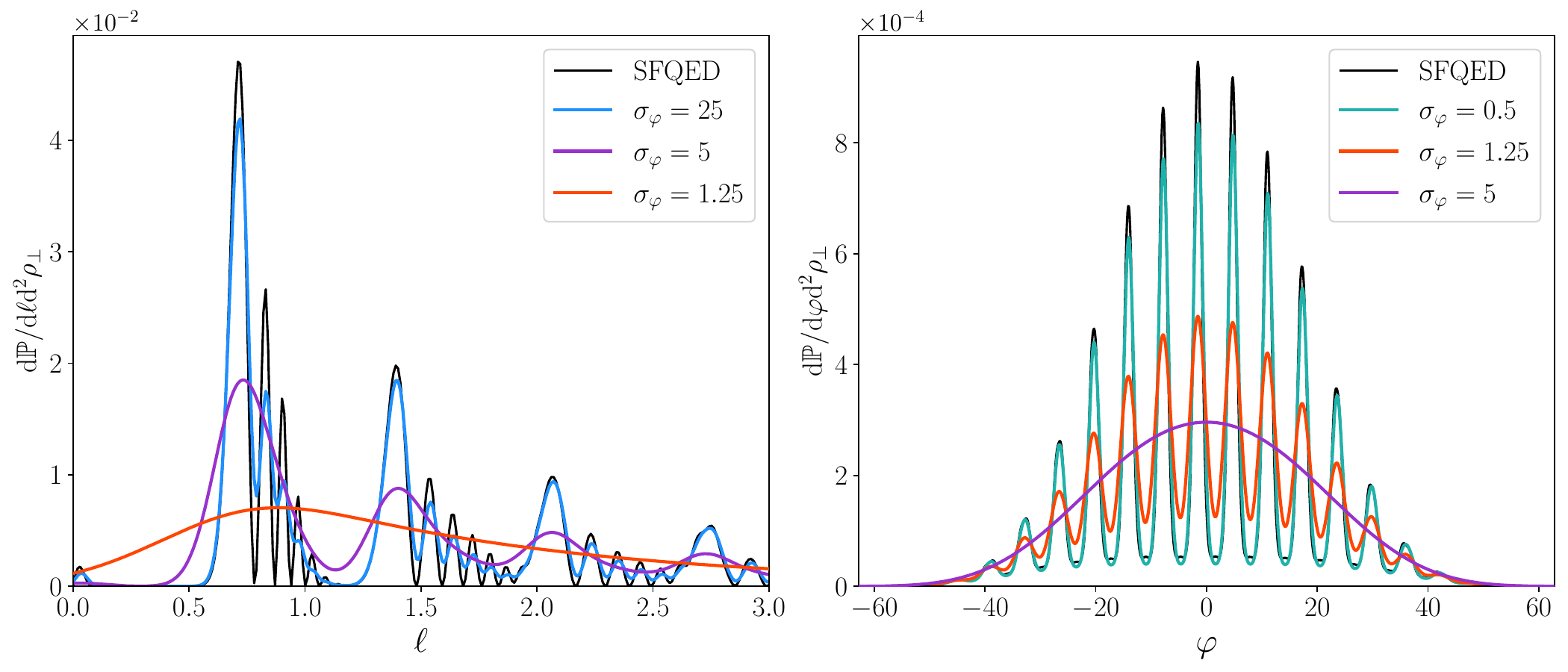}
\caption{Marginals of the Husimi JPD, Eqs.~\eqref{husimi_ell_marginal} and \eqref{husimi_phi_marginal} in the left and right panels, respectively, for various $\sigma_\varphi$.
They are compared with the `exact` marginals from SFQED without the Husimi transform, Eqs.~\eqref{energy_marginal} and \eqref{phase_marginal} (black curves).
The pulse profile and other parameters are the same as in Fig.~\ref{fig_Rate}.
    \label{fig_Husimi_marginals}}
\end{figure}

Also, since $\sigma_\varphi\sigma_\ell = 1/2$, we have only one independent parameter and in the following choose $\sigma_\varphi$ as the independent one; then $\sigma_\ell = 1/2\sigma_\varphi$. As we have seen, the fundamental time-frequency uncertainty prohibits us from simultaneous measurement of a photon's emission time and its energy with absolute precision, so we have to compromise via coarse-graining the JD. The Husimi transform allows us to achieve exactly this.
Selecting a specific value for $\sigma_\varphi$ establishes a particular choice in resolution: 
A small $\sigma_\varphi$ allows small uncertainty in the photon emission time distributions, but at the cost of large uncertainty of the energy of the emitted photon. Conversely, large values of $\sigma_\varphi$ permit high energy resolution at the cost of a large uncertainty when the interaction took place. This behaviour can be seen best in the marginal distributions
Eqs.~\eqref{husimi_ell_marginal} and \eqref{husimi_phi_marginal}. They are plotted in the left and right panels of Fig.~\ref{fig_Husimi_marginals}, respectively. Decreasing the value of $\sigma_\varphi$ in the left panel of Fig.~\ref{fig_Husimi_marginals} smears out first the fine sub-harmonic peaks, and for very short phase averaging intervals $\sigma_\varphi=1.25$ also the harmonic structure. The reverse situation happens in the right panel of Fig.~\ref{fig_Husimi_marginals}, where the strongly peaked structure of the probability rate as a function of phase gets smeared out more for longer phase averaging intervals.

\subsection{Husimi JPD vs local approximations}
\label{subsec_Husimi_vs_local}
To investigate how well the Husimi JPD \eqref{Husimi_new} can serve as a probability rate for NCS we now compare it to limiting cases where approximate probability rates are known. These two cases are the locally constant field approximation (LCFA) \cite{Ritus1985} and the locally monochromatic approximation (LMA) \cite{Heinzl2020, Larin2025}.

\subsubsection{Locally constant field approximation}
We consider the angular resolved LCFA probability rate \cite{Blackburn2020} and compare it with the angular resolved Husimi JPD, Eq.~\eqref{Husimi_new}. The former is based on the idea that the formation length of the emitted photon \cite{BAIER2005} is much smaller than the spatial and temporal inhomogeneities of the external field \cite{Ritus1985, DiPiazza2018}; thus, the field can be treated locally as a constant. In a plane-wave background, this is usually true when the conditions $a_0\gg1$ and $a_0^2/\eta\gg1$ are fulfilled \cite{Dinu2016}. 

\begin{figure}[h]
\centering
\includegraphics[width=\textwidth]{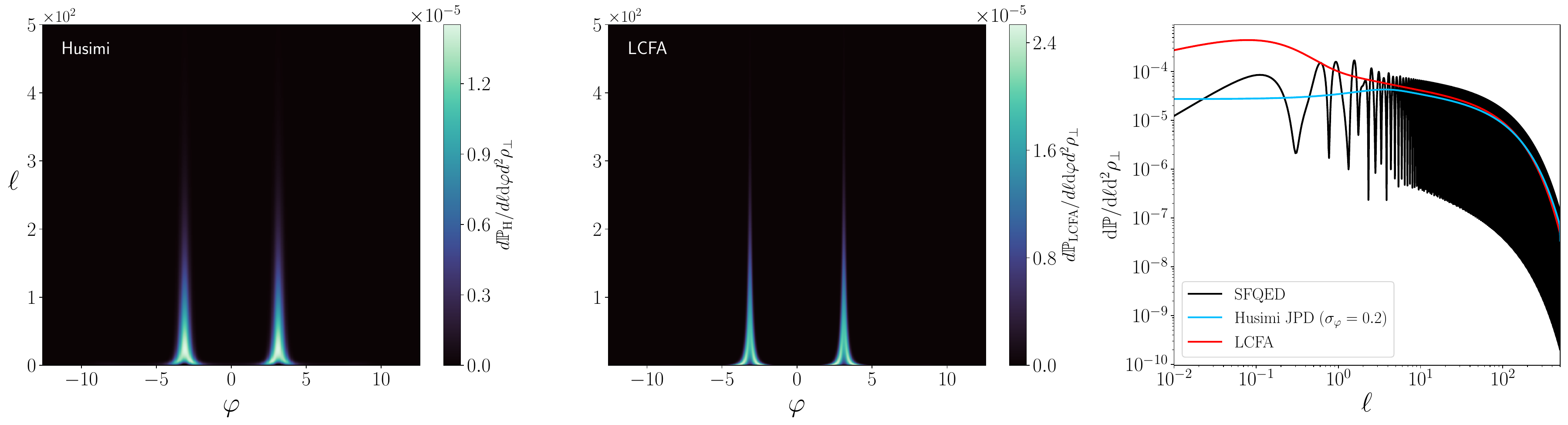}
\includegraphics[width=\textwidth]{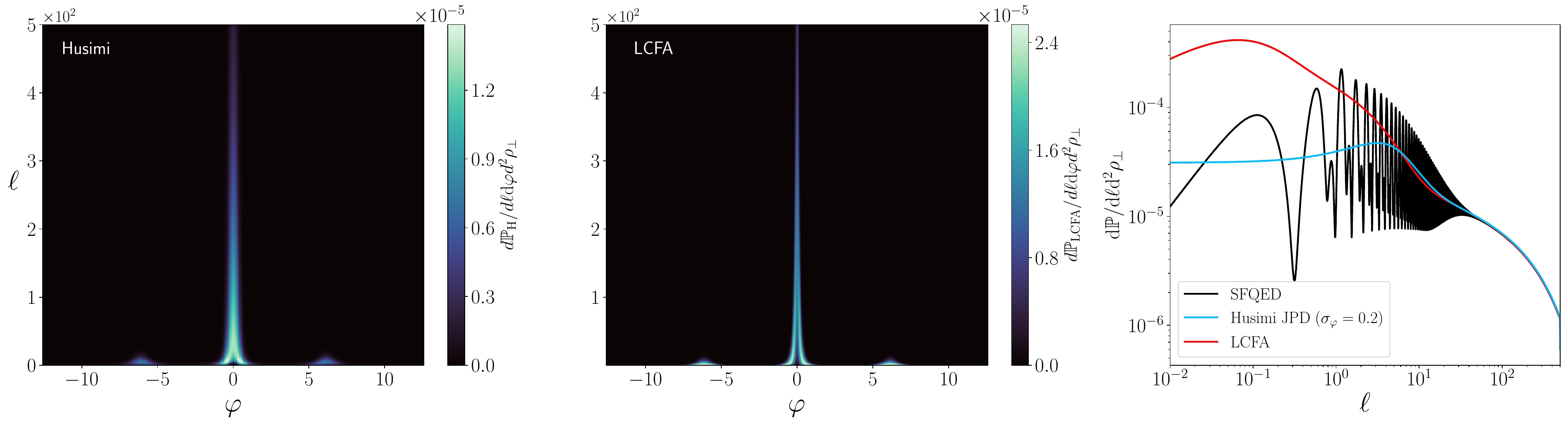}
\caption{Joint probability distributions for photon emission given by the Husimi JPD \eqref{Husimi_new} with $\sigma_\varphi=0.2$ (left column) and LCFA (central column). The right column shows the spectrum given by Husimi JPD compared with the spectra obtained with LCFA and exact SFQED calculations. The pulse profile is the same as for Fig.~\ref{fig_Rate} with $N=4$ and the other parameters are: $a_0=7$, $\vec{\rho}_\bot=\left(0,7\right)$ (top row) and $\vec{\rho}_\bot=\left(0,-7\right)$ (bottom row). 
    \label{fig_Husimi_vs_LCFA}}
\end{figure}

In Fig.~\ref{fig_Husimi_vs_LCFA} we demonstrate fully-differential probability rate for LCFA and Husimi JPD with relatively small $\sigma_\varphi = 0.2$, which corresponds to the original JD \eqref{Proto_rate} coarse-grained with a narrow Gaussian in phase $\varphi$ and a broad one in the energy variable $\ell$. Thus, we expect to see the phase structure to be resolved in great details, while the spectrum is smeared broadly. Also, in accordance with the LCFA assumptions, we expect a similar picture for the probability rate, since it depends on the local value of the field. Indeed, what we find is that both distributions are sensitive to the temporal structure of the laser pulse and predict smooth spectra, averaging through the oscillating curve, obtained by means of the numerical calculations of Eq.~\eqref{explicit_protorate}. While predictions of the LCFA and Husimi JPD agree in the high-energy part of the spectrum, the latter better approximates low-energy part of the full SFQED calculation. The LCFA overestimates the yield of photons with transferred momentum $\ell \lesssim (1 + \vec{\rho}_\bot^2)/a_0^2$ \cite{DiPiazza2018, Blackburn2020}; the cumulative effect of these low-energy photons manifests as a divergence in the angular integrated LCFA spectrum \cite{Ilderton2019}.

In the case of LCFA, the smooth structure of the spectrum is due to the fact that it is obtained by neglecting inhomogeneities of the external field and corresponding interference effects. For the Husimi JPD, however; the spectrum is smooth, because we coarse-grained the oscillating spectrum with Gaussian in the energy variable $\ell$ with effectively large $\sigma_\ell$. 
The advantage of the Husimi JPD over LCFA is, that it may be applied also outside the realm of applicability of the LCFA \cite{Dinu2016, DiPiazza2018}, see below.

Another shared feature of the Husimi JPD and LCFA probability rate is that for a given observation direction, $\vec k/k^0$, they both predict distinct radiation bursts at certain parts of the laser pulse. Depending on the observation direction, we see one (bottom row in Fig.~\ref{fig_Husimi_vs_LCFA}) or two (top row in Fig.~\ref{fig_Husimi_vs_LCFA}) consecutive emission bursts. Their number and position may be explained from the semiclassical picture \cite{Blackburn2018}. The ultrarelativistic electron propagating through an intense laser pulse emits along its instantaneous velocity (similar to the synchrotron radiation \cite{Ternov1995}), and the emission pattern in the transverse plane imprints the structure of the vector potential \cite{Blackburn2020}. If we choose the observation point, through which electron traverses once, we see one distinct peak in JPD; if it traverses twice, we see two peaks. Given all aforementioned properties, we conclude that the predictions of the LCFA probability rate and Husumi JPD for relatively small $\sigma_\varphi$ are in good agreement. 

\subsubsection{Locally monochromatic approximation}

Now as we studied relation between the LCFA and Husimi JPD, we turn to the locally monochromatic approximation and try to understand what are the differences and similarities between it and the Husimi JPD. The idea behind the LMA differs from the LCFA. The former relies on the separation of time scales in plane-wave-like backgrounds and subsequent averaging over the fast component \cite{Heinzl2020, Larin2025}. Such separation is possible only for slowly varying backgrounds, where the change of the envelope is negligible on the cycle scale. As a result of the cycle-averaging, LMA is insensitive to the cycle scale  structure of the pulse, but captures harmonic nature of the spectrum relatively well. We expect to see similar behavior for the Husimi JPD when we increase $\sigma_\varphi$, since such regime corresponds to the coarse-graining with wide Gaussian in $\varphi$ variable and narrow one in $\ell$ variable. We work with angular-resolved quantities and the standard angular-resolved LMA is known to be divergent, thus we choose to compare the Husimi JPD with the extension of LMA, the LMA$^+$ \cite{Larin2025}.

\begin{figure}[h]
\centering
\includegraphics[width=\textwidth]{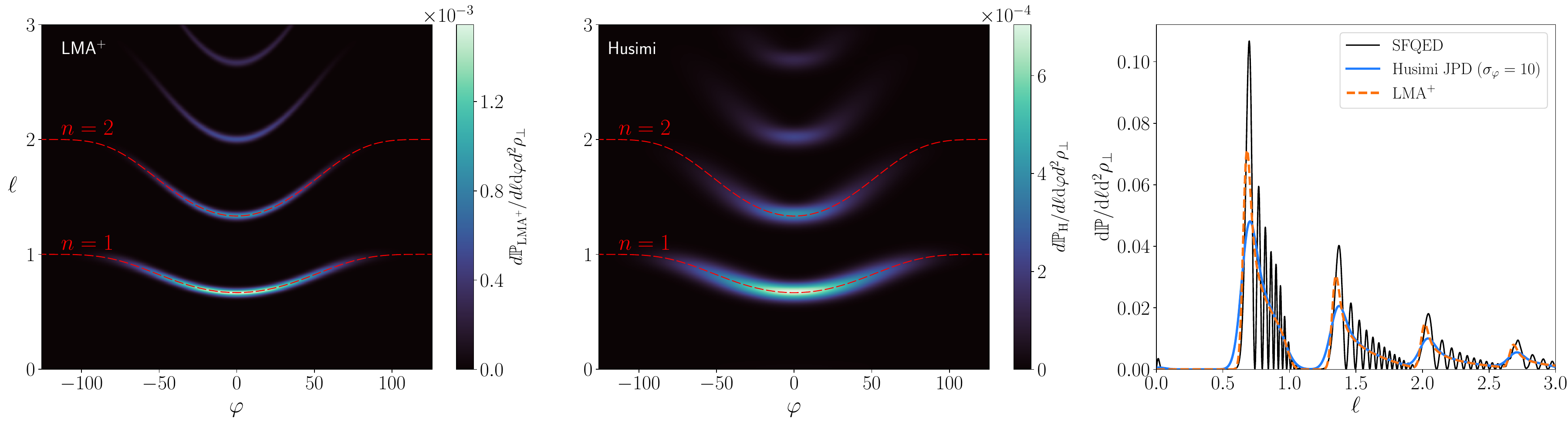}
\includegraphics[width=\textwidth]{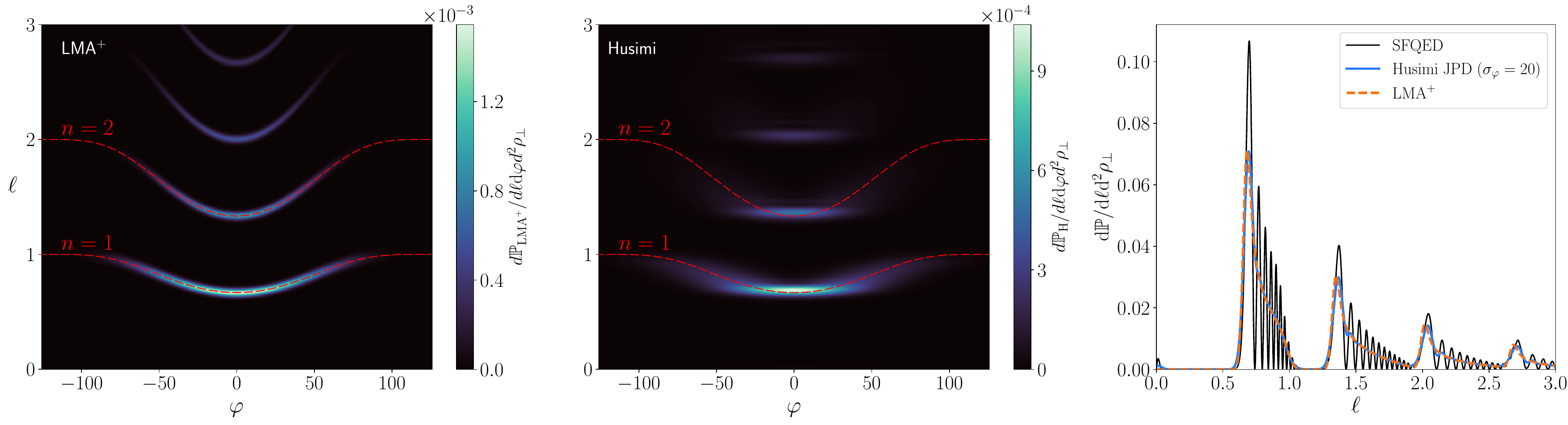}
\includegraphics[width=\textwidth]{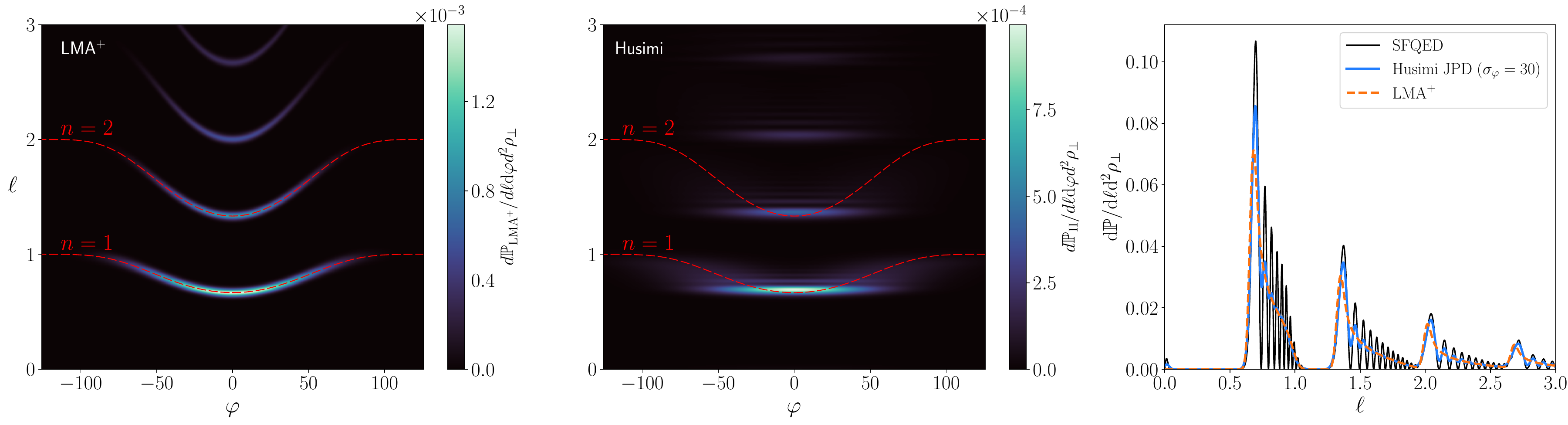}
\caption{Joint probability distributions for photon emission given by the LMA$^+$ (left column) and the Husimi transform \eqref{Husimi_new} (central column). The right column shows the spectra given by the Husimi JPD, compared with the spectra obtained with LMA$^+$ and exact SFQED calculations. The different rows show Husimi JPD for $\sigma_\varphi = 10$ (top), $\sigma_\varphi = 20$ (middle) and $\sigma_\varphi = 30$ (bottom). The dashed red lines depict the positions of the first and second harmonics. The number of cycles $N=40$, the potential and all other parameters are the same as in Fig.~\ref{fig_Rate}.
    \label{fig_Husimi_vs_LMA}}
\end{figure}

We compare the LMA$^+$ with the Husimi JPD for several relatively large values of $\sigma_\varphi$. In the left and central panels of the top row of Fig.~\ref{fig_Husimi_vs_LMA} we see that Husimi JPD for $\sigma_\varphi = 10$ and LMA$^+$ predict emission from the regions that follow theoretical prediction for the harmonics position: $\ell_n(\varphi) = n(1+\vec{\rho}_\bot^2)/(1 + \vec{\rho}_\bot^2 + \vec{a}^2_\mathrm{rms})$ \cite{Seipt2013} (red dashed lines in Fig.~\ref{fig_Husimi_vs_LMA}), reproducing their ponderomotive red-shift. In both distributions the half-integer harmonics disappeared (see, for comparison Fig.~\ref{fig_Rate}). For LMA$^+$ it happens because of the averaging over cycle scale \cite{Larin2025} and in Husimi JPD it is due to the convolution \eqref{Husimi_new}, which averages neighboring positive and negative values to zero.

The energy spectra, obtained from LMA$^+$ and Husimi JPD for $\sigma_\varphi = 10$ qualitatively similar, but do not agree exactly (see, right panel in top row in Fig.~\ref{fig_Husimi_vs_LMA}). We note that LMA$^+$ spectrum is always the same for the given pulse length $N$, however corresponding spectrum from Husimi JPD can be tuned by varying $\sigma_\varphi$. In the middle row in Fig.~\ref{fig_Husimi_vs_LMA} we demonstrate Husimi JPD and its $\ell$-marginal for $\sigma_\varphi = 20$, when the latter coincides with the LMA$^+$ spectrum. Increasing $\sigma_\varphi$ further we even may resolve sub-harmonic structure (see, right panel in bottom row in Fig.~\ref{fig_Husimi_vs_LMA}). However, the Husimi JPD starts to differ significantly from the LMA$^+$ result.

The convolution for larger $\sigma_\varphi$ betters resolution in $\ell$ variable, and we can resolve the sub-harmonic structure in Husimi JPD. But it coarse-grains the phase structure completely, so even the slowly-varying envelope component is indistinguishable anymore. The smooth harmonics  turns into the series of straight lines, confined between the low-energy ($\ell_n(0)$) and high-energy ($\ell_n(\pm\infty)$) edges of $n^\mathrm{th}$-harmonic.

\subsection{Optimal resolution}

Now that we have demonstrated that the Husimi joint probability distributions agrees well with the known local approximations 
for very short $\sigma_\varphi$ (low energy resolution like in the LCFA) and for long $\sigma_\varphi$ (good spectral resolution, but average over many laser cycles just as in the LMA),
we want to investigate if Husimi JPD can be of use in the intermediate range where both the LCFA and the LMA are not applicable. We argue that in the intermediate range there must be some value of $\sigma_\varphi$ for which both the (harmonic) spectrum and temporal (laser cycle) structures are optimally resolved.

We choose this value by optimizing the resolution for the aspect ratio of the $(\varphi,\ell)$-plane. Since the aspect ratio of the Husimi JPD is defined by the ratio of Gaussian dispersions, fixing it provides an additional condition that, along with $\sigma_\ell\sigma_\varphi=1/2$, allows us to pick $\sigma_\varphi$ uniquely. Given that characteristic variation distance for the laser phase is $2\pi$ and that separation distance for neighboring harmonics is of the order of unity, we want aspect ratio to be $1:2\pi$, thus we choose $\sigma_\ell/\sigma_\varphi=1/2\pi$. The optimal value of the smearing parameter, preserving structure in both canonically conjugate variables then becomes $\sigma_\varphi=\sqrt{\pi}$.

 \begin{figure}[h]
\centering
\includegraphics[width=0.6\textwidth]{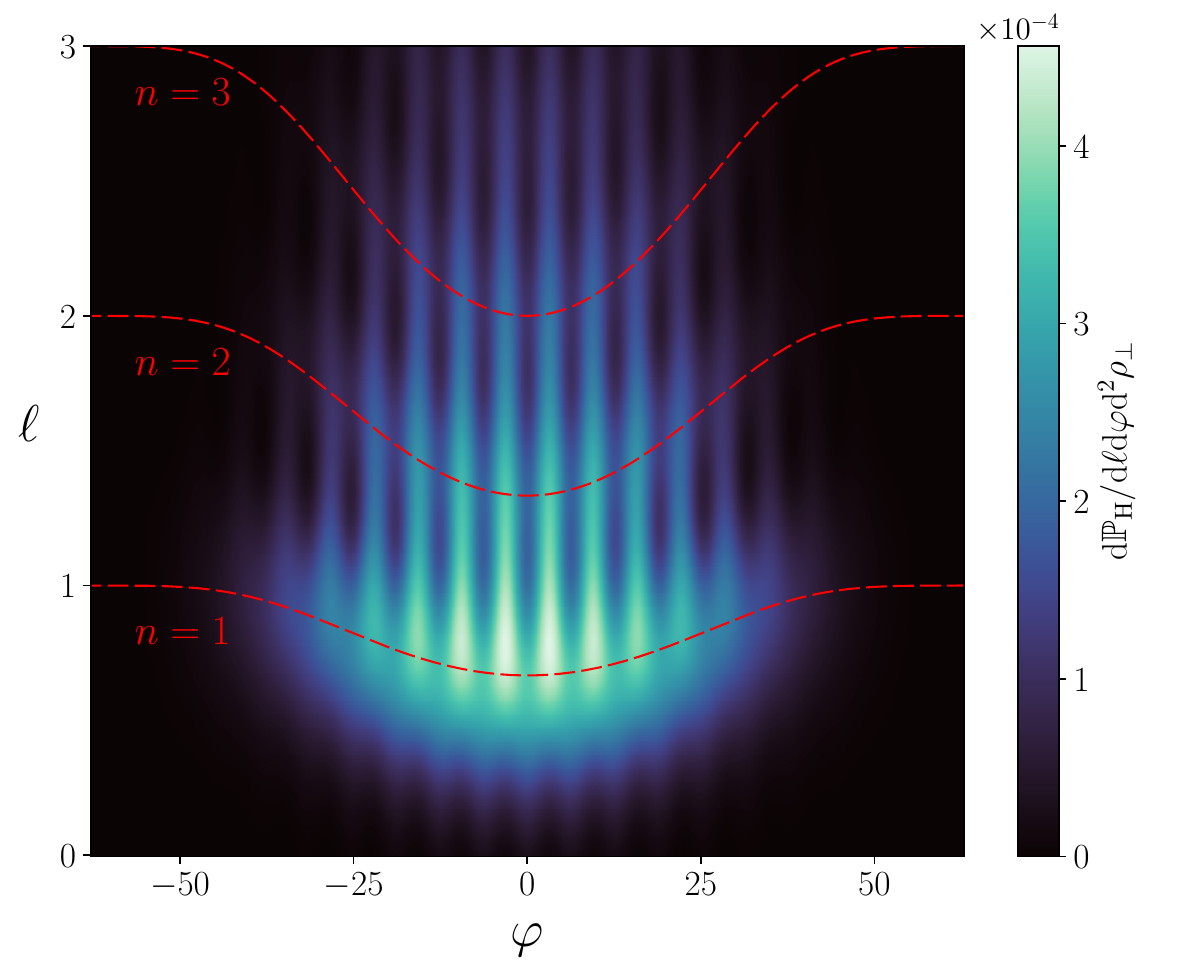}
\caption{Husimi JPD for photon emission with $\sigma_\varphi=\sqrt{\pi}$. The positions of the first three harmonics are marked with red dashed lines. The parameters and laser pulse profile are chosen as in Fig.~\ref{fig_Rate}. 
    \label{fig_Husimi_optimal}}
\end{figure}

In Fig.~\ref{fig_Husimi_optimal} we plot optimally smeared Husimi JPD for the photon emission. Comparing it with the original joint distribution Fig.~\ref{fig_Rate}, we see how the positions of the harmonics and their ponderomotive red-shift are reproduced correctly. The half-integer harmonics are absent, but Husimi JPD still predicts radiation from the region between the harmonics, which follows the cycle structure of the pulse. Thus, Husimi JPD in the intermediate regime ($\sigma_\varphi = \sqrt\pi$) is capable of capturing features of both LCFA and LMA simultaneously.

\section{Results and Application} 
\label{sec_results}
The temporal structure of the external laser pulse directly affects the spectrum and angular distribution of the emitted radiation. By tailoring the laser pulse, one can control key features of the emitted radiation \cite{Krajewska_2014, Shao:PRA2023, Krafft:PRAB2023,  Timoshenko:PRA2025}. In this section, we use the Husimi joint probability distribution to study how complex pulse configurations, such as pulses with different carrier-envelope phase or with variable polarization, influence photon emission. We show that the Husimi JPD provides a clear and intuitive picture of the emission process, linking spectral features to their temporal origin.

\subsection{Carrier-envelope phase effects}
\label{subsec_CEP}

\begin{figure}[h!]
\centering
\includegraphics[width=\textwidth]{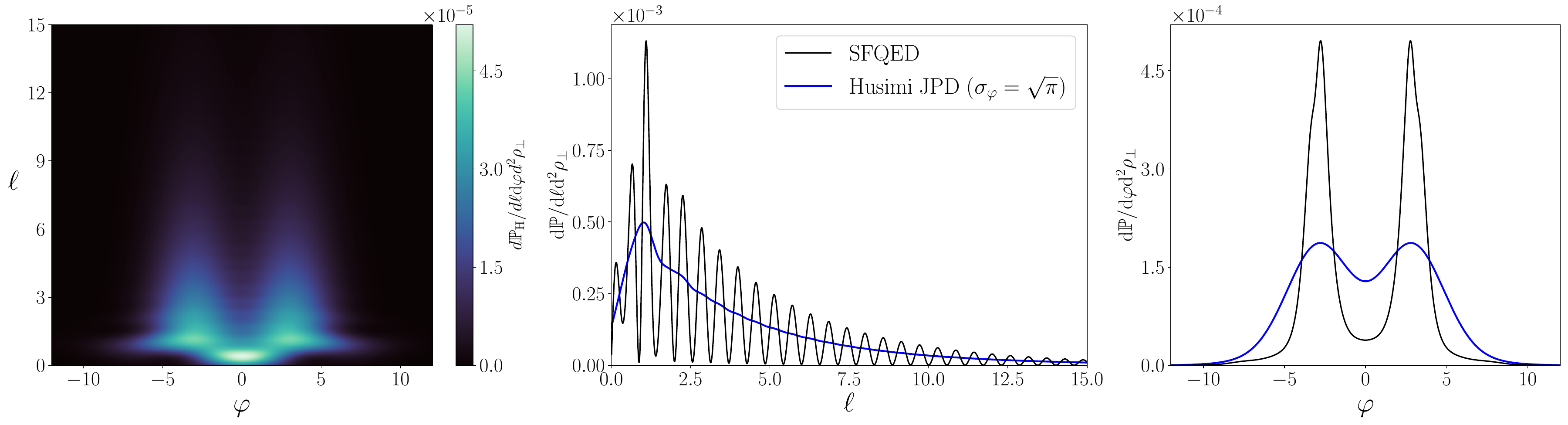}
\includegraphics[width=\textwidth]{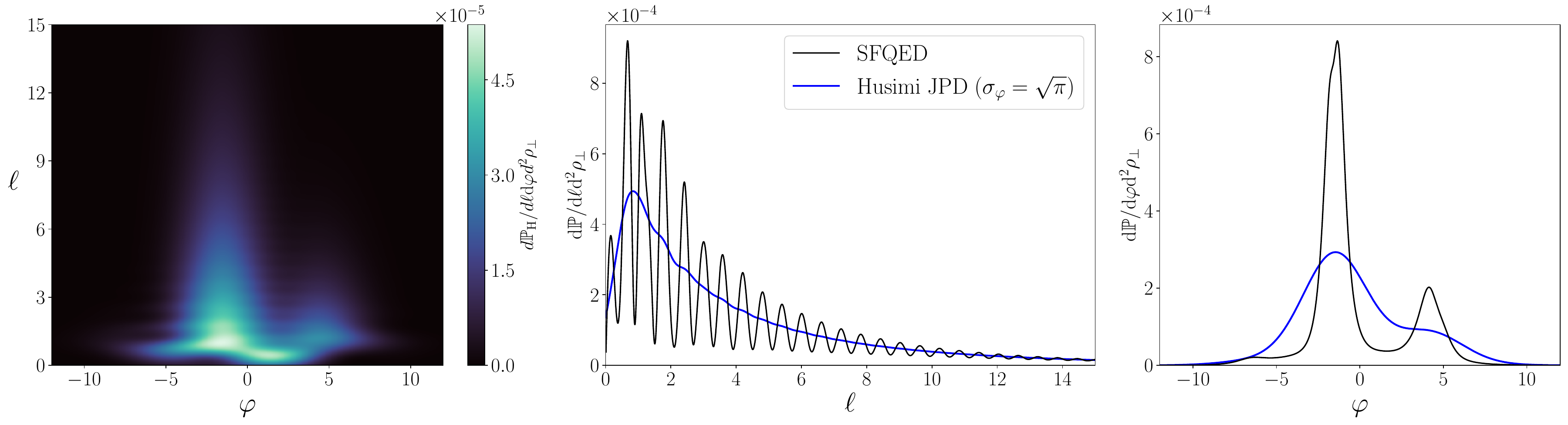}
\caption{{Husimi JPD for photon emission and its marginal distributions compared with the exact SFQED results. The laser profile is ${{\vec{f}}_ \bot }\left( \phi  \right) =  \exp\left(-\phi^2/2N^2\right)\left( {\sin \left(\phi + \phi_\mathrm{CEP}\right) , \cos \left(\phi + \phi_\mathrm{CEP}\right) } \right)$ with $N = 4$ and carrier-envelope phase $\phi_\mathrm{CEP} = \pi/2$ (top row) and $\phi_\mathrm{CEP} = 0$ (bottom row). The other parameters are: $a_0 = 2$, $\vec{\rho}_\bot = \left(a_0, 0\right)$, $\eta = 0.1$.}
    \label{fig_CEP}}
\end{figure}

\begin{figure}[h!]
\centering
\includegraphics[width=\textwidth]{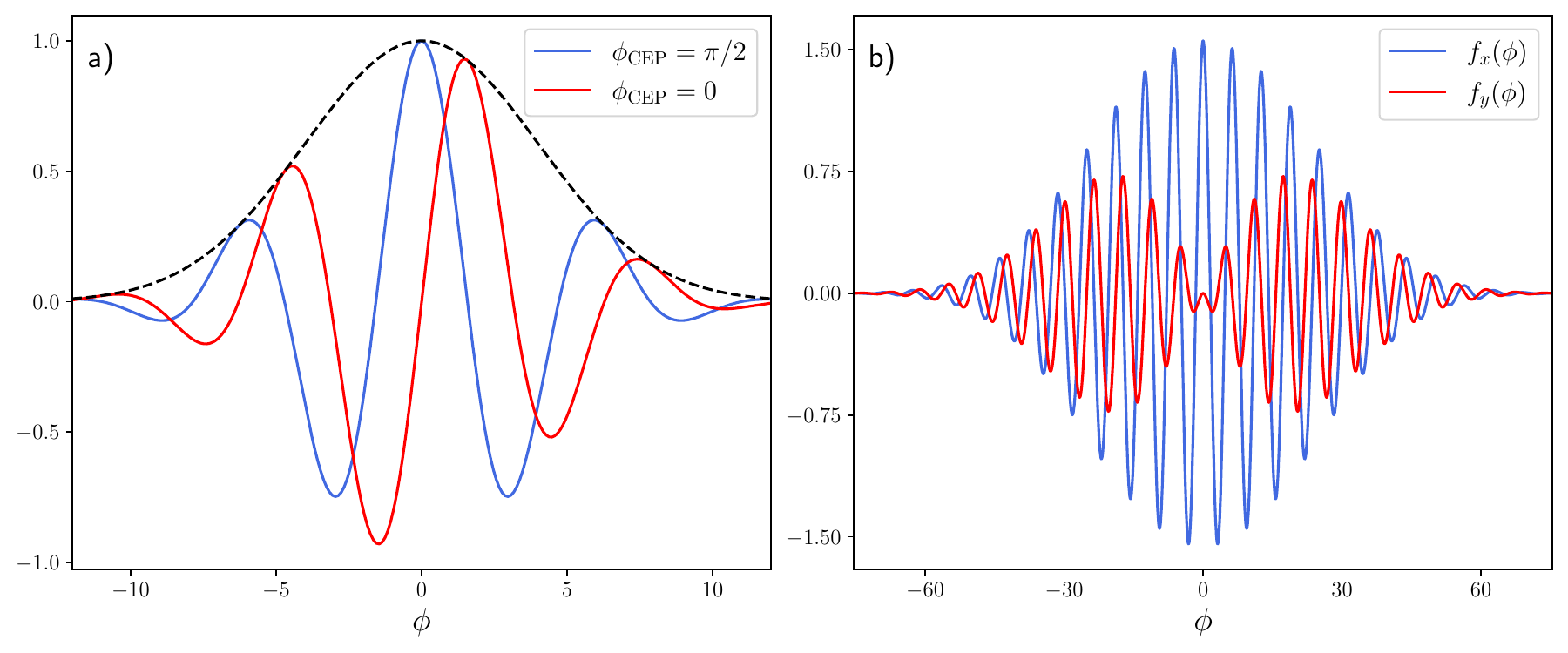}
\caption{a) the $f_x(\phi)$ components of the vector potential with different carrier envelope phases ($\phi_\mathrm{CEP}=\pi/2$ - blue solid line and $\phi_\mathrm{CEP}=0$ - red solid line) and their envelope (black dashed line), b) the components of the vector potential with varying polarization ($f_x(\phi)$ - blue solid line and $f_y(\phi)$ - red solid line).
    \label{fig_potential}}
\end{figure}

The spectrum of emitted radiation \cite{Boca2009} and its angular distribution \cite{Seipt2013a} depend on the relative carrier-envelope phase (CEP) of the laser pulse. The shorter the pulse, the more pronounced influence of the CEP. In Fig.~\ref{fig_CEP} we plot two Husimi JPDs of photon emission (for the optimal phase window $\sigma_\varphi=\sqrt{\pi}$) and their marginals for the two pulses with different CEP. The features of JPD may be explained if we consider the profile of the vector potential (see, Fig.~\ref{fig_potential}a). According to the semi-classical picture, the radiation from the ultrarelativistic electron is beamed along its instantaneous velocity and in transverse plane imprints shape of the vector potential $\rho_\bot = -\vec{a}_\bot(\varphi)$ \cite{Blackburn2020}. In the left and right panels of the top row in Fig.~\ref{fig_CEP} we see how for $\rho_\bot=(a_0,0)$ the emission comes from the two neighboring peaks of the $\cos$-carrier, approximately located at $\varphi =\pm\pi$. In turn, the left and right panels of the bottom row shows that for $\sin$-carrier the emission at the given observation point comes dominantly from a single peak, roughly at $\varphi=-\pi/2$.

The ability of the Husimi JPD to capture cycle-scale structures and CEP-dependent signatures in photon emission makes it a potential tool for pulse characterization. Currently, most proposals for determining the CEP of ultra-intense pulses rely on measuring asymmetries in the angular distribution of radiation produced in electron-laser collisions \cite{Mackenroth2010, Titov:PRD2016, Li:PRL2018}. However, recent breakthroughs in generating high-intensity zeptosecond X-ray pulses \cite{Chen2025, Li2025} suggest an alternative path forward: by employing these pulses in a streaking technique \cite{IPP2011}, the Husimi JPD for photon emission can be employed for a direct time-frequency mapping of electron dynamics, providing a robust determination of the carrier-envelope phase.

\subsection{Application to Polarization gating}
\label{subsec_PG}

To further demonstrate the capabilities of the Husimi JPD we now consider NCS in the laser pulse with varying polarization. Such pulses are produced by means of the polarization gating (PG) technique \cite{Corkum1994}. This method has, for instance, been used to generate intense single cycle attosecond pulses from high-harmonic generation on plasma surfaces \cite{Yeung:PRL2015}. Only recently it was suggested to use PG for obtaining narrowband gamma radiation \cite{Valialshchikov2021} in SFQED. Here we apply our Husimi JPD to NCS within PG pulses to gain further insights into the generation of the gamma radiation for such complex laser configuration.

The pulse with varying polarization may be realized as an overlap of two co-propagating circularly polarized pulses with opposite helicities. Here, we adapt the expression for the potential from \cite{Valialshchikov2021}:
\begin{align}
f_x(\phi)=g\left(\phi + \frac{\delta}{2}\right)\cos\left(\phi+\frac{\delta}{2}\right) + g\left(\phi - \frac{\delta}{2}\right)\cos\left(\phi-\frac{\delta}{2}\right),\\
f_y(\phi)=g\left(\phi + \frac{\delta}{2}\right)\sin\left(\phi+\frac{\delta}{2}\right) - g\left(\phi - \frac{\delta}{2}\right)\sin\left(\phi-\frac{\delta}{2}\right),
    \label{PG_pulse}
\end{align}
with $g(\phi) = \exp(-\phi^2/N^2)$, and choose the same parameters as in \cite{Valialshchikov2021}: $N = 8\pi$ and $\delta=N$ (delay between two pulses). Such choice corresponds to the pulse with the polarization that changes from the circular to the linear at the pulse center and back.

\begin{figure}[h!]
\centering
\includegraphics[width=\textwidth]{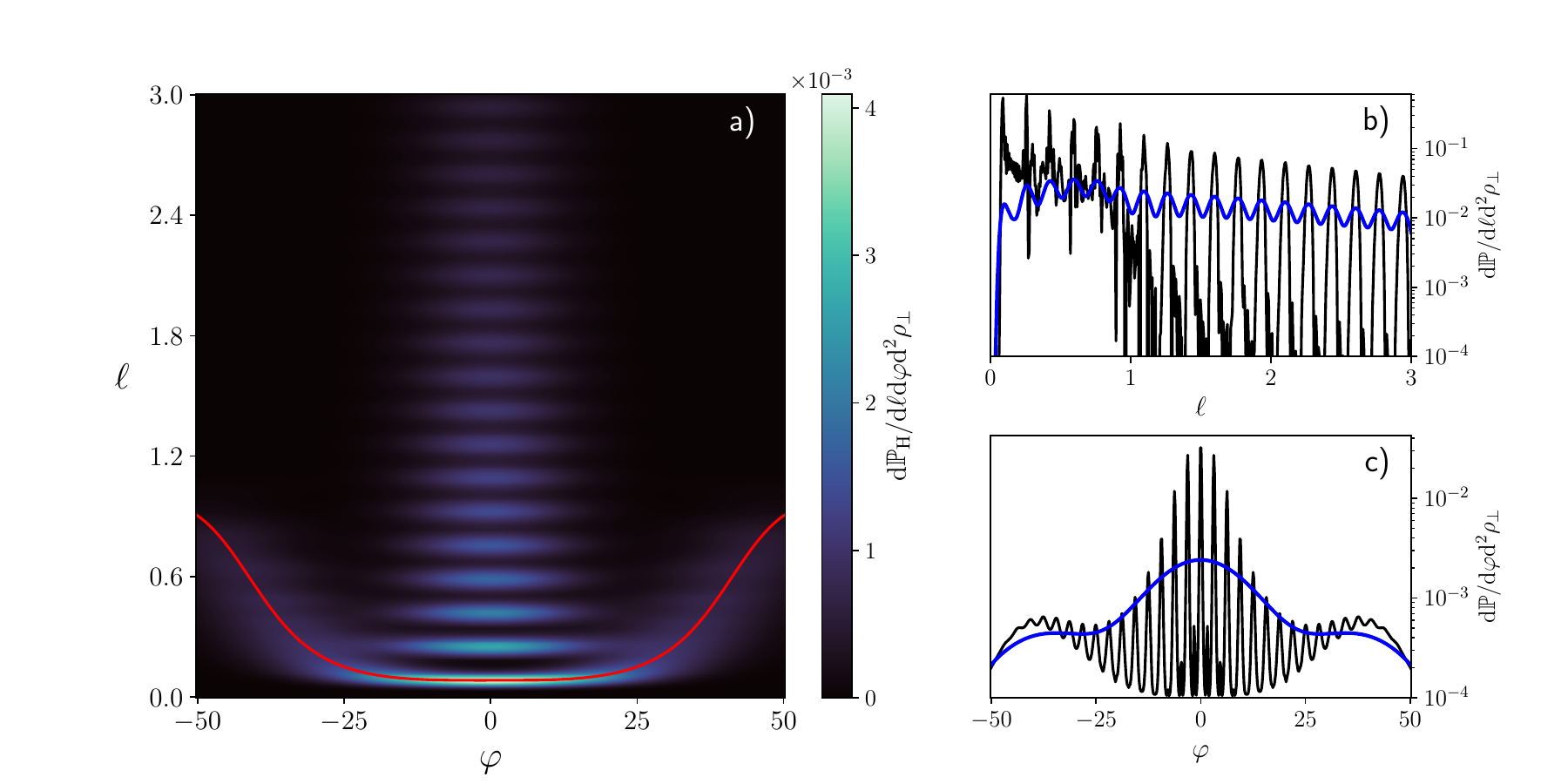}
\caption{{Husimi JPD of photon emission within the pulse with phase-dependant polarization ($\sigma_\varphi=10$). The parameters are $a_0=3$, $\vec{\rho}_\bot=\left(0,0\right)$, for the pulse parameters see the text. }
    \label{fig_PG}}
\end{figure}

The Husimi JPD in Fig.~\ref{fig_PG} perfectly demonstrates the essence of the PG technique as a method for generation of narrow bandwidth gamma radiation. The higher harmonics emitted only in the vicinity of the pulse peak, where the polarization of the background field is close to linear (see, Fig.~\ref{fig_potential}b). Otherwise, their emission is prohibited and only fundamental harmonic contributes to the spectrum \cite{Harvey2009}. The harmonics emitted close to the laser envelope maximum are weakly influenced by the ponderomotive broadening. On contrary, we can see how the fundamental harmonic is significantly broadened due to the phase-dependant red-shift (see red line in Fig.~\ref{fig_PG}a). Moreover, it overlaps with some higher harmonics and leads to the broad spectrum in the low-energy part of the spectrum (see, Fig.~\ref{fig_PG}b).

The possible applications of the Husimi JPD are not limited to the discussed two examples. Many other pulse configurations (e.g., chirped pulses \cite{Seipt2015}, multicolor setups \cite{Narozhny_multicolor}, train pulses \cite{Krajewska_2014})  may be studied by means of the developed approach. Another intriguing application is studying of the classical \cite{DiPiazza2021} and quantum \cite{Podszus2022} radiation reaction. Since both effect lead to the suppression of the emitted spectrum, we expect to see their influence in Husimi JPD. We assume that some parts of the laser pulse will contribute less due to the classical or quantum radiation reaction and propose Husimi JPD as a tool to study, where exactly these changes take place. Thus, analysis based on the Husimi JPB may help to clarify the mechanisms behind particle energy loss. 

\section{Conclusion}
\label{sec_summary}

In this article we showed how the concept of a joint distribution in the time-frequency domain can be utilized within the SFQED framework. Starting from the $S$-matrix formalism, we derived the JD by mod-squaring the photon emission amplitude. Instead of evaluating the two phase integrals, we retained explicit dependence on the average laser phase and only performed integration over the interference window. This approach naturally accounts for interference effects and allows their influence on the emitted spectra to be observed from a temporal perspective. 
While the resulting JD reproduces the correct marginal distributions, in particular the correct emission spectrum,
the JD itself alternates in sign, preventing a straightforward probabilistic interpretation of the time-frequency behaviour.

To overcome this limitation, we adapted methods from time–frequency analysis and derived a Husimi joint probability distribution for photon emission, in which the original JD is coarse-grained with correlated Gaussian kernels in $\varphi$ and $\ell$. The resulting Husimi JPD is strictly non-negative and can be interpreted as the probability of emitting a photon with a given energy at a specific laser phase. Its marginal distributions correspond to the original JD marginals smoothed by the corresponding Gaussian function. 

We studied the properties of the Husimi JPD and showed how its resolution can be tuned to explore different regimes. The resolution is controlled by a single parameter $\sigma_\varphi$. The detailed energy spectra with coarse-grained phase information is achieved for relatively small values of $\sigma_\varphi$, whereas the detailed phase distributions with smeared energy spectra correspond to the larger $\sigma_\varphi$. We showed that in an intermediate regime both the features in the time and frequency domain remain visible. We compared the Husimi JPD with two widely used local approximations (LCFA and LMA) that provide non-negative probability rates. Husimi JPD for relatively small $\sigma_\varphi$ shares a lot of common features with LCFA, in turn for the larger $\sigma_\varphi$ its predictions coincide with the extended LMA (LMA$^+$). By controlling coarse-graining
of the Husimi JPD, we may interpolate between the LCFA and LMA regimes without being constrained by their typical applicability criteria, thereby effectively bridging the gap between these two approximation methods.

Finally, we applied the Husimi JPD to study various tailored laser pulse configurations, including pulses with different carrier-envelope phases and pulses with variable polarization throughout the pulse (polarization gating). In all cases, the Husimi JPD predictions agree with expectations and allow to gain further insight into the temporal evolution. Thus, the Husimi JPD may serve as a tool in designing complex laser pulses to generate radiation with desired spectral properties. 
Our results suggest that the Husimi JPD could serve as a valuable tool for investigating not only nonlinear Compton scattering, but also other SFQED processes like Breit-Wheeler pair production, as well as phenomena like classical and quantum radiation reaction.

\bibliography{article_lib}

\end{document}